\documentclass[fleqn,12pt,twoside]{article}
\usepackage{bbm}
\usepackage{espcrc1}

\usepackage{graphicx}
\usepackage[figuresright]{rotating}

\newcommand{\pom}{\mathbbm{P}}

\title{Diffraction, the Color Glass Condensate and String Theory}

\author{Thomas Bittig\address{Institut f\"ur Theoretische Physik, 
        Universit\"at Heidelberg, \\ 
        Philosophenweg 16, D-69120 Heidelberg, Germany}%
        \thanks{Email: T.Bittig@thphys.uni-heidelberg.de}, 
        Carlo Ewerz\address{Dipartimento di Fisica, Universit{\`a} di Milano, 
        and INFN,\\
        Via Celoria 16, I-20133 Milano, Italy}%
        \thanks{Email: Carlo.Ewerz@mi.infn.it}}
       
\begin{document}

\maketitle

\begin{abstract}
We explain the main ideas of the color glass condensate
in high energy collisions.
Different approaches to the problem are outlined 
with emphasis on the resummation approach. We present
evidence that the color glass condensate can be described
by an effective conformal field theory or even by a string
theory.
\end{abstract}

\section{DIFFRACTION AND THE COLOR GLASS CONDENSATE}
\label{sec:cgc}

Hadronic scattering processes at high energies are called diffractive if 
there is a wide angular region in the final state without hadronic activity. 
Such a rapidity gap can only emerge if a colorless object, the Pomeron, 
is exchanged between the colliding particles. 
Clearly, elastic scattering at high energy is a diffractive process. Hence 
the very same object determines via the optical theorem also the total 
cross section at high energies. Due to Regge or high energy factorization 
the elastic amplitude is a convolution of the amplitude for Pomeron 
exchange with impact factors coupling it to the external particles, 
and the actual energy dependence is encoded only in the Pomeron. 

Although there is plenty of data for different scattering processes, 
we still have no conclusive picture of the Pomeron in terms of 
quarks and gluons, 
and to understand the high energy limit of QCD remains one of 
the most challenging open problems in the physics of the strong interaction. 
There are two key points which make the problem complicated. 
Many scattering processes at high energies involve low momentum 
scales, in general prohibiting a perturbative approach. Fortunately though, 
there are some scattering processes involving momentum scales 
sufficiently large for making perturbation theory applicable, like 
for example the scattering of two highly virtual photons 
which split into quark-antiquark pairs and then scatter off each 
other. But even if one concentrates on these latter processes there 
is still a second problem, namely the occurrence of high parton densities. 
At high energies there is a large phase space for the emission of 
soft partons off the incoming hadrons. Eventually, the usual 
picture breaks down in which different parton cascades in the 
hadron are viewed as being independent of each other. At this 
point the partons in the hadron start to overlap and recombination 
effects have to be taken into account. As a consequence, the evolution 
of the system with energy becomes nonlinear, asking for new 
theoretical methods to describe it. An advantage of high parton 
densities, on the other hand, is that multiple 
scattering in a dense medium can induce hard scales, possibly 
extending the applicability of perturbation theory to a larger 
number of scattering processes. 

The dense system of partons in colliding hadrons at high energies 
consists mainly of gluons, because these are the particles of highest 
spin and largest color charge. Since gluons carry a color charge and are 
bosons, one calls this dense parton system the color glass condensate (CGC). 
The `glass' refers to the fact that the quantum evolution of the system, 
that is the emissions and recombinations of gluons, takes place 
on much longer time scales than the interaction in the 
collision process. In other words, the hadron evolves for a long time 
before the actual collision takes place. 

\section{THEORETICAL DESCRIPTION OF THE COLORED GLASS}

The color glass condensate has been studied in a number of different 
approaches in perturbative QCD. The approach on which we will 
concentrate in the present talk is based on the concept of resummation 
of large logarithms of the energy $\sqrt{s}$ which can compensate 
the smallness of the coupling constant. In the leading logarithmic 
approximation (LLA) \cite{Kuraev:fs,Balitsky:ic} one collects all terms 
of the form $(\alpha_s \log s)^n$ in the perturbative series. 
The resulting linear evolution equation, the BFKL equation, describes 
the Pomeron as an exchange of two interacting reggeized gluons in 
the $t$-channel. 
At higher energies it becomes necessary to go beyond the LLA by taking 
into account nonlinearities, as we will describe below. 
Among the other theoretical approaches to the color glass condensate 
are the dipole 
picture \cite{Mueller:1993rr}, the operator expansion of Wilson lines 
\cite{Balitsky:1995ub}, and the 
`color glass condensate approach'\footnote{Somewhat confusingly, 
that approach has been given the same name as the object it 
describes.}, for 
a review see \cite{Iancu:2003xm}. In all these approaches one finds 
identical or at least similar results for the main properties of the color 
glass condensate. One of them is the fact that the main dynamics takes place 
in the two-dimensional transverse space of the scattering process. In the 
approximation of low parton densities, in which nonlinear terms can be 
neglected, all the approaches reproduce the BFKL Pomeron. 
We are convinced that a more detailed understanding of the relations 
between the different approaches will be crucial for future 
developments in the theory of the CGC. 

In the resummation approach one can include high density effects 
and go beyond the LLA by taking into account exchanges with 
more than two gluons in the $t$-channel. Here one collects the 
maximally possible number of logarithms for a given number 
of exchanged gluons, giving rise to the generalized 
leading logarithmic approximation (GLLA). This can be done either 
by keeping the number of gluons fixed during the $t$-channel 
evolution \cite{Bartels:1978fc,Bartels:1980pe,Kwiecinski:1980wb}, 
or, more interestingly, in the extended form of that approximation 
(EGLLA) in which the number of $t$-channel gluons is allowed to 
fluctuate \cite{Bartels:unp,Bartels:1999aw}. In the EGLLA, amplitudes 
for the production of $n$ gluons in the $t$-channel are described 
by a hierarchy of coupled integral equations. It is important to note 
that in the EGLLA it is not necessary to take the large-$N_c$ limit 
at any stage. 

The analysis \cite{Bartels:1993ih,Bartels:1994jj,Bartels:1999aw} 
of the amplitudes in the EGLLA has shown that they can be cast 
into the form of an effective field theory of reggeized gluons. In 
this effective theory only states with (fixed) even numbers of gluons 
occur which are coupled to each other by effective vertices 
$V_{2 \to 2l}$. 
So far the two-to-four \cite{Bartels:1994jj} and the two-to-six 
gluon vertex \cite{Bartels:1999aw} have been calculated explicitly. 
The emerging structure is that of a two-dimensional field theory 
in transverse space, with rapidity $Y \sim \log s$ as an additional 
time-like parameter. 
Due to lack of space we mention only in passing here that the 
field theory structure emerges because of the reggeization of the gluon 
at high energies \cite{Lipatov:1976zz,Ewerz:2001fb}. In short, reggeization 
reflects the fact that at high energies the correct degrees of freedom are 
collective excitations of the Yang-Mills field rather than elementary 
gluons. 

A very important property of the effective transition vertices 
is their conformal invariance in two-dimensional impact parameter 
space \cite{Bartels:1995kf,Ewerz:2001uq}, a property which they 
share with the kernel of the BFKL equation \cite{Lipatov:1985uk}. 
This observation makes it appear likely that the effective field theory of 
the color glass condensate in the EGLLA is in fact a conformal field 
theory (CFT). There is some hope that the known effective vertices are 
already sufficient to identify the underlying CFT. It would clearly 
be a significant step towards understanding the color glass condensate 
if one could apply the powerful methods of CFT to this problem. 
It could also help to establish a relation of high energy QCD to 
string theory via the AdS/CFT correspondence. 

A slightly less ambitious but still very interesting goal would be to 
look for an effective field theory of interacting Pomerons for the 
color glass condensate. Clearly, it is an additional approximation to 
restrict oneself to color singlet states of two gluons, and depending 
on the process under investigation it is not even necessarily a good one. 
For hard scattering processes on large nuclei it should be a valid 
approach though, since the coupling of the Pomerons to different 
nucleons makes these exchanges the leading ones in an expansion 
in $1/N_c$. In a first approximation one takes into account 
only the splitting of a Pomeron into two Pomerons, that is 
the three-Pomeron vertex, as the first nonlinear term in the 
evolution. In the limit of large $N_c$ this gives rise to the BK equation 
\cite{Balitsky:1995ub,Balitsky:1998ya,Kovchegov:1999yj,Kovchegov:1999ua} 
which is obtained in all the approaches to the CGC mentioned above 
in suitable approximations, and which is now widely used in the 
phenomenological study of high energy scattering processes. 
One clearly expects that with increasing parton densities also 
higher nonlinear terms and in general also $N_c$-suppressed 
terms become relevant. Obviously, 
it is very important for understanding the high energy limit of 
QCD to know whether higher Pomeron vertices exist, how 
they look like, and whether an effective field theory of interacting 
Pomerons can be formulated. In the following sections we will present 
new results \cite{BE2005} which bring us closer to 
an answer to these questions. Again, the conformal 
invariance plays a key role here. 

\section{CONFORMAL BOOTSTRAP FOR POMERON VERTICES}

Due to the conformal invariance of the BFKL equation in 
impact parameter space the resulting Pomeron states can be 
classified according to their behavior under $\mbox{SL}(2,\mathbbm{C})$ 
transformations. The representations of this symmetry group are 
characterized by the conformal weight $h = (1+n)/2 + i \nu$ with 
$n \in \mathbbm{Z}$ and $ \nu \in \mathbbm{R}$. 
Consequently, the Pomeron vertices depend on the quantum numbers $h_i$ 
of the Pomerons and on their coordinates $\rho_i$ in impact parameter 
space (understood as a complex plane). For simplicity we will restrict 
ourselves to Pomeron states with vanishing conformal spin $n_i=0$ 
in the following. 

In the resummation approach the three-Pomeron vertex $V_{3 \pom}$ 
is obtained from the effective two-to-four gluon vertex $V_{2\to 4}$ by 
projecting the four outgoing gluons onto a pair of BFKL Pomerons 
\cite{Lotter:1996vk,Korchemsky:1997fy}. 
The four-Pomeron vertex $V_{4\pom}$ has been calculated from the 
two-to-six gluon vertex $V_{2 \to 6}$ in a similar way in \cite{Ewerz:2003an}. 
Note that a four-Pomeron vertex has also been obtained in the approach based 
on the expansion of Wilson lines in \cite{Balitsky:2001mr}, but its relation to 
the vertex $V_{4\pom}$ has not yet been studied. 

The three-Pomeron vertex computed from the EGLLA consists of two 
terms which come with different powers of $N_c$, 
\begin{equation}
V_{3 \pom} (\rho_i, h_i) 
= g^4 C_1 
\left[V^{(0)}(\rho_i, h_i) + \frac{C_2}{N_c^2} \, 
 V^{(1)}(\rho_i, h_i) \right]
\end{equation}
where $C_{1,2}$ do not depend on the positions $\rho_i$. 
The corresponding four-Pomeron vertex 
has been obtained in \cite{Ewerz:2003an} as a sum of different 
terms which are all of the same order in $N_c$. These terms can 
be expressed in terms of three different functions 
$\Phi (\rho_j, h_j)$, $\Theta (\rho_j, h_j)$, and $\Pi(\rho_j, h_j)$ 
for which explicit integral representations exist. Interestingly, 
the vertex functions in $V_{3 \pom}$ and $V_{4\pom}$ are all completely 
symmetric in the Pomerons, despite the fact that they have been 
calculated from effective vertices which are not symmetric under the 
exchange of the two incoming and the four (resp.\ six) outgoing gluons. 
So far, the two vertices $V_{3 \pom}$ and $V_{4\pom}$ are the only 
Pomeron vertices which have been derived in the EGLLA, and a calculation 
of higher Pomeron vertices appears prohibitively complicated 
if the same techniques were to be used. As we will see, the properties 
of the two known vertices make it already possible to predict many 
properties of all higher vertices. 

A first important observation is that these two Pomeron vertices depend 
on the Pomeron coordinates in the particular form required by 
conformal symmetry, that is they are conformal three- and four-point 
functions, respectively. We emphasize that this is a nontrivial outcome 
of the EGLLA and has not been put in. Further, an interpretation of 
these vertices in the framework of an effective CFT of interacting 
Pomerons would require that the four-point function is related to 
the three-point function in the form of a conformal bootstrap relation. 
Remarkably, relations of exactly this type can be found \cite{BE2005}. 
They express the functions $\Phi$, $\Theta$ and $\Pi$ appearing 
in the four-Pomeron vertex as products of those appearing in the 
three-Pomeron vertex, $V^{(0)}$ and $V^{(1)}$. 
For example, the function $\Phi(\rho_j, h_j)$ ($j=a,b,c,d$) 
appearing in the four-Pomeron vertex $V_{4 \pom}$ can be expressed 
as 
\begin{equation}
 \Phi = \sum_{n_k=-\infty}^{+\infty}\, \int_{-\infty}^{+\infty} d\nu_k \,
f(h_k) \int d^2\rho_k \, 
V^{(1)}(\rho_a^*, \rho_b^*, \rho_k, h_a^*, h_b^*, h_k)
\, V^{(1)}(\rho_k^*, \rho_c^*, \rho_d, h_k^*,h_c^*,h_d) 
\end{equation}
with a weight factor $f(h_k) = |h_k -1/2|^2/\pi^4$. This result 
is derived with the help of a completeness relation and it 
is therefore crucial to sum over a full set of intermediate Pomeron 
states with label $k$. 
Similarly, one can express the other two functions $\Theta$ and 
$\Pi$ in the four-Pomeron vertex as products of the two 
functions $V^{(0)}$ and $V^{(1)}$, where for $\Pi$ this product 
has to be taken with crossed arguments. The product of 
$V^{(0)}$ with itself does not occur. 
A closer inspection shows that it cannot be obtained 
in the four-Pomeron vertex $V_{4 \pom}$ derived in the EGLLA because the 
Feynman diagrams required for it are not part of that approximation. 
The latter observation might help in identifying possible important 
contributions beyond the EGLLA, and to compare them for example 
with those of \cite{Peschanski:1997yx}. 

These bootstrap relations between the three-Pomeron vertex and the 
four-Pomeron vertex strongly suggest that there exists in fact an 
effective CFT of interacting Pomerons. Assuming that they are not 
simply accidental, one can use them to formulate conjectures 
about higher $n$-Pomeron vertices for arbitrary $n$. Our analysis 
of the two known Pomeron vertices makes it hence possible 
to predict -- up to normalization factors -- all higher Pomeron vertices 
as they would be obtained in the EGLLA. Explicit formulae will 
be given in \cite{BE2005}. 

\section{STRING AMPLITUDES FOR POMERON VERTICES}

It has long been conjectured that QCD at high energies is a string 
theory or can at least be related to one. In fact high energy hadron 
scattering is even the origin of string theory. It is therefore interesting 
to see whether the known interactions of Pomerons resemble 
string amplitudes. A relation of this kind was discussed in the context 
of a conjecture for higher Pomeron vertices in the dipole picture in 
\cite{Peschanski:1997yx}. We have found that the Pomeron vertices 
obtained in the EGLLA do not coincide with those conjectured 
there. Nevertheless, it turns out that the Pomeron vertices in the 
EGLLA can be related to string amplitudes in a way similar to that 
proposed in \cite{Peschanski:1997yx}. 

Specifically, we find that both $V_{3 \pom}$ and $V_{4 \pom}$ can be 
expressed as integrands of Virasoro-Shapiro amplitudes of a closed 
bosonic string theory \cite{BE2005}. 
Writing for example $V^{(1)}$ as a conformal three-point 
function (again assuming vanishing conformal spins $n_i$), 
\begin{equation}
V^{(1)} = \Lambda(\nu_a,\nu_b,\nu_c) \!\! 
\prod_{i<j \atop i,j \in \{a,b,c\}} 
\!\!\!\!\! |\rho_{ij}|^{-2\Delta_{ij}} 
\end{equation}
with $\Delta_{ab}=h_a+h_b-h_c$ and $\rho_{ij}= \rho_i - \rho_j$, 
it can be shown that 
\begin{equation}
A_6(p_a,p_b,p_c,p_\delta,p_1,p_2) = 
\int d^2\rho_\delta \, \left| \frac{\rho_{ca}}
{\rho_{c\delta} \rho_{\delta a}} \right|^2 \, 
\Lambda(\nu_a,\nu_b,\nu_c) 
\end{equation}
is a Virasoro-Shapiro amplitude for the scattering of six closed string 
tachyon states 
after suitable identification of the string momenta with the scaling 
dimensions $\nu_i$ of the Pomerons. More precisely, we have to 
identify the scalar products of the string momenta in the target 
space with combinations of the scaling dimensions. Similar relations have 
been found for the four-Pomeron vertex, and one can also find them 
for the higher Pomeron vertices which follow from our conjecture in the 
previous section. Intriguingly, that correspondence to string amplitudes 
holds also for vertices of Pomerons with nonvanishing conformal spins 
which can be related to amplitudes of excited string states. 

At present, there are still many open questions concerning the 
relation of Pomeron vertices to string amplitudes. It turns out, 
for example, that the string amplitude for a given $n$-Pomeron vertex 
is not unique. Instead, there is only a minimal number of strings 
required for each given $n$. Due to that it is possible that the 
number of strings matches the number of gluons, but an identification 
of closed strings with Pomerons seems to be excluded. 
Other key problems are to interpret the Pomeron quantum 
numbers via string momenta in a suitable Minkowskian target space, 
and to find the meaning of 
the critical dimension of closed string theory in the context of Pomeron 
vertices. Finding an effective string theory for interacting Pomerons 
would open many interesting possibilities, including the computation 
of phenomenologically relevant Pomeron loop amplitudes. 

\section{SUMMARY}

We have studied the color glass condensate in the approach based 
on the perturbative resummation of logarithms of the energy. 
The Pomeron vertices obtained in this approach 
exhibit bootstrap relations and hint at an underlying effective 
conformal field theory of interacting Pomerons. We find 
some evidence for the exciting possibility that the color glass 
condensate can even be described by an effective string theory. 

\section*{ACKNOWLEDGMENTS}

C.\,E.\ was supported by a Feodor Lynen fellowship of the 
Alexander von Humboldt Foundation.

\end{document}